\documentclass[review,12pt,sort&compress,1p]{elsarticle}

\usepackage{epsfig}

\usepackage{amssymb,amsmath,amsthm}
\usepackage{amsfonts}
\usepackage{graphicx}

\usepackage{amssymb}
\usepackage{lineno,hyperref}

\usepackage{dcolumn}% Align table columns on decimal point
\usepackage{bm}% bold math
\usepackage{float}
\usepackage{subfigure}
\usepackage{graphics}
\usepackage{array}
\usepackage{color}

\journal{Fluid Phase Equilibria}

\begin{document}

\begin{frontmatter}



\title{Construction of a minimum energy path for the VT flash model by an exponential time differencing scheme with the string method}


\author[1]{Yuze Zhang}
\ead{1906391168@pku.edu.cn}

\author[2]{Yiteng Li}
\ead{yiteng.li@kaust.edu.sa}

\author[1]{Lei Zhang}
\ead{zhangl@math.pku.edu.cn}

\author[2]{Shuyu Sun\corref{cor1}}
\ead{shuyu.sun@kaust.edu.sa}

\cortext[cor1]{Corresponding author: Shuyu Sun.}

\address[1]{Beijing International Center for Mathematical Research, Peking University, Beijing 100871, China}
\address[2]{Physical Science and Engineering Division (PSE), King Abdullah University of Science and Technology (KAUST), Thuwal 23955-6900, Saudi Arabia}

\begin{abstract}
Phase equilibrium calculation, also known as flash calculation, plays significant roles in various aspects of petroleum and chemical industries. Since Michelsen proposed his milestone studies in 1982, through several decades of development, the current research interest on flash calculation has been shifted from accuracy to efficiency, but the ultimate goal remains the same focusing on estimation of the equilibrium phase amounts and phase compositions under the given variable specification. However, finding the transition route and its related saddle points are very often helpful to study the evolution of phase change and partition. 
Motivated by this, in this study we apply the string method to find the minimum energy paths and saddle points information of a single-component VT flash model with the Peng-Robinson equation of state. As the system has strong stiffness, common ordinary differential equation solvers have their limitations. To overcome these issues, a Rosenbrock-type exponential time differencing scheme is employed to reduce the computational difficulty caused by the high stiffness of the investigated system.
%Numerical examples are provided to show the feasibility and accuracy of the given algorithm.
In comparison with the published results and experimental data, the proposed numerical algorithm not only shows good feasibility and accuracy on phase equilibrium calculation, but also successfully calculates the minimum energy path and and saddle point of the single-component VT flash model with strong stiffness. 
\end{abstract}


% \begin{highlights}
% \end{highlights}

\begin{keyword}
VT flash \sep Peng-Robinson equation of state \sep minimum energy path \sep string method \sep Rosenbrock-type exponential time differencing scheme



\end{keyword}

\end{frontmatter}


\section{Introduction}

% Modeling and simulating multi-phase fluid systems with real equation of state (EOS) (such as Peng-Robinson equation of state \cite{PR}) has being a popular and challenging research topic in reservoir engineering, especially for the phase equilibrium calculation problem which often appears in many complex phase behaviors. 
%%%介绍一下背景%%
The accurate knowledge of phase equilibria is of vital importance in petroleum industry to determine the number of equilibrium phases and their amounts and compositions for complex reservoir fluids, which stimulates the development of equation-of-state-based phase equilibrium calculation. As one of important applications, phase equilibrium calculation is an integral component of compositional multiphase flow simulators that are used to model recovery processes sensitive to compositional changes, such as primary depletion of volatile oil and gas condensate reservoirs and multiple contact miscible flooding \cite{Chen2007}. Moreover, it can be used as standalone calculation as well for analyzing, designing and optimizing processes and facilities 
in order to reduce undesired species (e.g. $\text{H}_{2}\text{O}$) for conforming to product specifications, remove acid gas (e.g. $\text{CO}_{2}$ and $\text{H}_{2}\text{S}$) for protecting pipelines and equipment from corrosion \cite{Poormohammadian2015,Wang2017,Wang2018}, determine the amount of inhibitor (e.g. methanol and glycols) for avoiding gas hydrate formation \cite{Anderson1986,Pedersen1996,Haghighi2009,Burgass2018}, control asphaltene precipitation for enhancing flow assurance \cite{Sabbagh2006,Shirani2012,Jindrova2016}, etc.

The most commonly-used phase equilibrium calculation is conducted at the specified pressure ($P$), temperature ($T$) and chemical composition ($\bm{N}$), which is known as PT flash calculation in the petroleum industry. Usually, the Wilson correlation is used to initialize flash calculation but this approximation may not converge to the equilibrium solution. To overcome this issue, stability test can be used to determine whether phase split takes place and it will give a better initial approximation if the investigated fluid is unstable at the given condition but requires additional computational cost.
The milestone studies of Michelsen in stability testing \cite{Michelsen1982a} and phase splitting calculation \cite{Michelsen1982b} laid a foundation for the development of PT flash calculation using a single consistent equation of state (EOS), such as the Soave-Redlich-Kwong (SRK) EOS \cite{SRK} and the Peng-Robinson (PR) EOS \cite{PR}. 

Recently, the computation of phase equilibria under the given volume ($V$), temperature ($T$) and mole composition ($\bm{N}$), also known as VT flash calculation, has become a promising alternative of the conventional PT flash. In addition 
to its well-posed formulation, VT flash exhibits an unambiguous relation between pressure and volume so that the root selection procedure in PT flash framework can be avoided \cite{Jindrova2013}. Without the preprocessing of successive substitution iteration, the second-order VT flash methods exhibit higher computational efficiency than the second-order PT flash methods in stability testing. Also, it can improve the overall efficiency of phase split calculation while the performance somewhat depends on the investigated systems and thermodynamic models \cite{Liang2018}. Another important advantage of VT flash formulation is phase behavior modeling for unconventional hydrocarbon mixtures in the presence of capillary pressure \cite{Sandoval2019,Yiteng2018,Sun2019,Nichita2019}, since the commonly-used interfacial tension models are explicit function of molar density rather than pressure. Numerous efforts have been made in the past few years to expand the applications of VT flash not only in various phase equilibrium problems \cite{Travalloni2014,Luo2018,Yiteng2020} but also in compositional flow simulation \cite{Polivka2014,Paterson2018}.

Clearly, all the aforementioned phase equilibrium calculation focuses on the estimation of phase amounts and phase compositions at the equilibrium state where either the Gibbs free energy or Helmholtz free energy is minimized at the end of PT or VT flash calculation respectively. However, finding transition routes between 
% different equilibrium states 
different local minima, one of which actually represents the equilibrium state, 
under certain conditions is often helpful to study the evolution of phase partitioning. These transition routes can be called as minimum energy paths (MEPs), where the potential force is always parallel to the path. In addition, we can get the first-order saddle points' information of the original energy landscape of the investigated system, which are bottlenecks for a large number of particular barrier-crossing events. Furthermore, the knowledge of high-order saddle points is helpful in finding multiple local minima \cite{hiosd1,hiosd2,hiosd3}, as well as the relative minimum point of energy stability, and even the global minimum point can be obtained through enumeration. 
% This has certain directive effect to the petroleum industry concretely. 
These knowledge could help phase behavior modeling of complex reservoir fluids.
In this study, we focus on calculating the MEP and first-order saddle points of a single-component VT flash model. There are several numerical schemes proposed to calculate the MEPs \cite{WeinanE2007,MEP1,MEP2,MEP3,MEP4,MEP5}. Among them, the string method \cite{WeinanE2007,MEP4} exhibits a good performance and enables to efficiently determine energy transition pathways for complex systems with smooth energy landscape by evolving strings rather than points in configuration space. This approach consists of two main steps, evolution of the string by some ordinary differential equation (ODE) solvers and reparametrization of the string by interpolation. The accuracy of this approach significantly relies on ODE solvers. It is worth mentioning that equation-of-state-based VT flash models could have ill-conditioned Hessian matrices with very large condition numbers 
(For example, in this study PR EOS is employed to model a single-component system, the two eigenvalues of the Hessian matrix differ by $10^{8}-10^{9}$ times). 
Moreover, explicit numerical approaches are required because the real scale of the problem is huge. Some common ODE solvers like Euler and Runge-Kutta methods cannot converge or require a very small time step. For these problems with strong stiffness, an exponential time differencing (ETD) scheme is used here to reduce the influence of the rigid term during numerical calculation.

The ETD scheme involves exact integration of the governing equations followed by an explicit approximation of a temporal integral for the nonlinear terms. It can provide satisfactory accuracy and stability even for problems with strong stiffness. This scheme was originally applied in the field of computational electrodynamics \cite{ETD5}. Then it was systematically studied in \cite{ETD1} and was developed for solving stiff systems by Cox and Matthews \cite{ETD2}. Also, in \cite{ETD2}, high-order multi-step ETD schemes and Rounge-Kutta versions of ETD schemes were discussed. The linear stabilities of some ETD and modified ETD schemes were investigated by Du and Zhu \cite{ETD3,ETD4}. For some problems with very strong stiff terms, a bad chosen linearization of ordinary ETD schemes may cause a severe step size restriction. To avoid these problems, Hochbruck and Ostermann proposed Rosenbrock-type ETD scheme \cite{rosen1,rosen2} to handle the stiff nonlinear term without loss of the original stability. Thus, in this study, we use the Rosenbrock-Euler ETD scheme to solve the single-component VT flash model with strong stiffness. To the best of the authors' knowledge, this is the first time that the energy landscape of the VT flash problem is modeled by ETD schemes.
%In general, this scheme has a distinctive feature that it has exact evaluation of the contribution of the linear part, which can provide satisfactory accuracy and stability even for problems with strong rigidity. 
The structure of this paper is organized as follows. In Section 2, we describe the VT flash model using PR EOS and give details for the single-component system. In Section 3, the framework of string method is provided. In Section 4, we establish the first-order and second-order ETD schemes for the single-component VT flash problem. In Section 5, numerical examples are presented to demonstrate the performance of the proposed ETD schemes and construct the MEP for the investigated system. At the end, we make some remarks and conclusions.

\section{VT flash calculation}

Let us consider a reservoir fluid mixture of $M$ components occupies volume $V$ at temperature $T$. If the investigated mixture is unstable and splits into at most two phases, the total Helmholtz free energy of a two-phase system is defined by
\begin{equation} \label{eq:HelmholtzFreeEnergy}
	F = F_{1}(\bm{N}_{1},V_{1},T) + F_{2}(\bm{N}_{2},V_{2},T)\, ,
\end{equation}
where $\bm{N}_{\alpha} = \left[N_{1,\alpha}, \ldots, N_{M,\alpha}\right]^{T}$ and $V_{\alpha}$ denote the mole number vector and volume of phase $\alpha = 1, \, 2$, respectively. 
For brevity, we will drop $T$ in the following expressions of $F$ since $T$ is assumed to be constant. Given that the total mole number of each component ($N_{i}$) and total volume ($V$) are fixed, we can arbitrarily select the mole composition and volume of one phase as primary variables and compute the counterparts of the other phase by the mole and volume balance equations
\begin{align}
    N_{i,1} + N_{i,2} &= N_{i} \, , \quad i = 1,\ldots,M \\
    V_{1} + V_{2} &= V \, .
\end{align}

The second law of thermodynamics indicates the minimum principle of Helmholtz free energy at given $\bm{N}$, $V$ and $T$ so that the phase equilibrium problems can be formulated as the following minimization problem \cite{Firoozabadi2015} if $\bm{N}_{1}$ and $V_{1}$ are chosen as the primary variables
\begin{equation} \label{eq:ConstrainMin}
    \min F\left(\bm{N}_{1},V_{1}\right)
\end{equation}
subject to 
\begin{subequations}
    \begin{eqnarray}
        0 \leq N_{i,1} \leq N_{i} \, , \label{eq:constraint1}\\
        V_{1} - \sum_{i=1}^{M}b_{i}N_{i,1} > 0 \, , \label{eq:constraint2}\\
        (V-V_{1}) - \sum_{i=1}^{M}b_{i}\left(N_{i}-N_{i,1}\right) > 0 \, , \label{eq:constraint3}
    \end{eqnarray}
\end{subequations}
where $b_{i}$ is the co-volume parameter of pure component. For details, the parameters of PR EOS can be found in \ref{sec:appA}. In terms of the relation $F = -PV + \sum_{i}\mu_{i}N_{i}$, the partial derivatives of $F(\bm{N}_{1},V_{1})$ with respect to $N_{i,1}$ and $V_{1}$ yield
\begin{align}
	\frac{\partial F}{\partial N_{i,1}} &= \frac{\partial F_{1}\left(\bm{N}_{1}, V_{1}\right)}{\partial N_{i,1}} + \frac{\partial F_{2}\left(\bm{N}_{2}, V_{2}\right)}{\partial N_{i,1}} = \mu_{i,1} - \mu_{i,2} \label{eq:ChemicalEquilibrium} \, , \\
	\frac{\partial F}{\partial V_{1}} &= \frac{\partial F_{1}\left(\bm{N}_{1}, V_{1}\right)}{\partial V_{1}} + \frac{\partial F_{2}\left(\bm{N}_{2}, V_{2}\right)}{\partial V_{1}} = P_{2} - P_{1} \, , \label{eq:MechanicalEquilibrium}
\end{align}
where $\mu_{i,\alpha}$ and $P_{\alpha}$ are the checmical potential of component $i$ and pressure in phase $\alpha$. Equation \eqref{eq:ChemicalEquilibrium} and \eqref{eq:MechanicalEquilibrium} are known as the chemical equilibrium condition and mechanical equilibrium condition, respectively. 
At the equilibrium state where $F$ is minimized, $\mu_{i,1} = \mu_{i,2}$ holds for all the components and meanwhile $P_{1} = P_{2}$ is achieved.

To efficiently solve bulk phase equilibria problems, one popular approach is making use of the local curvature of the energy function. The symmetric Hessian matrix has been extensively used to design fast and robust numerical algorithms for VT flash calculations \cite{Mikyska2012,Jindrova2013,Jindrova2015,Nichita2017,Nichita2018}. 
With a proper line search scheme, the original constrained minimization problem, comprising \eqref{eq:ConstrainMin} and \eqref{eq:constraint1}--\eqref{eq:constraint3}, can be reformulated as an unconstrained minimization problem and it can be solved as follows
\begin{equation}
	\mathbb{H}\Delta\bm{x} = -\bm{g} \, ,
\end{equation}
where
\begin{equation}
	\mathbb{H} = 
	\begin{bmatrix}
		\begin{array}{*{4}{c}}
			\cline{1-4}
			\multicolumn{1}{|c}{} &            				   & \multicolumn{1}{c|}{} & \multicolumn{1}{c|}{}  \\
			\multicolumn{1}{|c}{} & \bm{\mathbb{B}} & \multicolumn{1}{c|}{} & \multicolumn{1}{c|}{\bm{\mathbb{C}} }  \\
			\multicolumn{1}{|c}{} &            				   & \multicolumn{1}{c|}{} & \multicolumn{1}{c|}{}  \\
			\cline{1-4}
			\multicolumn{1}{|c}{} & \bm{\mathbb{C}}^{\text{T}} & \multicolumn{1}{c|}{} & \multicolumn{1}{c|}{\bm{\mathbb{D}} }  \\
			\cline{1-4}
		\end{array}
	\end{bmatrix} \, , \quad
	\Delta\bm{x} = 
	\begin{bmatrix}
		\Delta N_{i,1} \\
		\vdots \\
		\Delta N_{M,1} \\
		\Delta V_{1} \\
	\end{bmatrix} \, , \quad
	\bm{g} = 
	\begin{bmatrix}
		\mu_{i,1}  - \mu_{i,2}\\
		\vdots \\
		\mu_{M,1}   - \mu_{M,2}\\
		-P_{1}  + P_{2}\\
	\end{bmatrix} \, , 
\end{equation}
with
\begin{subequations}
    \begin{align}
        \mathbb{B}_{ij} &= \frac{\partial \mu_{i,1}}{\partial N_{j,1}} +  \frac{\partial \mu_{i,2}}{\partial N_{j,2}}\, ,\\
        \mathbb{C}_{j} &=  \frac{\partial \mu_{i,1}}{\partial V_{1}} +  \frac{\partial \mu_{i,2}}{\partial V_{2}} =  -\frac{\partial P_{1}}{\partial N_{i,1}} - \frac{\partial P_{2}}{\partial N_{i,2}} \, , \\
        \mathbb{D} &=  -\frac{\partial P_{1}}{\partial V_{1}} -  \frac{\partial P_{2}}{\partial V_{2}} \, .
    \end{align}
\end{subequations}
In order to ensure a constant decline of Helmholtz free energy over iterations, $\mathbb{H}$ is required to be positive definite as indicated in the aforementioned literature. Consequently, the positive definiteness of $\mathbb{H}$ is examined at each iteration for the sake of safety, and then a modified Cholesky factorization \cite{Gill1974,Schnabel1999} is performed if $\mathbb{H}$ is found not sufficiently positive definite. Clearly, such a process requires additional operations on the Hessian matrix. On the other hand, it has been reported in \cite{Paterson2018} that the modified Cholesky algorithm may hinder the convergence sometimes, although it is truly helpful in most instances. In comparison, the proposed numerical algorithm, which combines the ETD scheme and the string method, circumvents the modified Cholesky algorithm. 
For a single-component fluid system, $\mathbb{H}$ is a $2\times 2$ matrix. Considering the fluid is a pure substance, we will drop the subscript $i$ that is commonly used as component index. Then the chemical potential and pressure of pure component reads as
\begin{equation} 
    \resizebox{\textwidth}{!}{%
		$\begin{aligned}
			\mu = - RT\ln\left(\frac{V-B}{N}\right) + RT\frac{B}{V-B} + 	 \frac{A}{2\sqrt{2}BN}\ln\left(\frac{V+(1-\sqrt{2})B}{V+(1+\sqrt{2})B}\right) - \frac{aNV}{V^{2}+2BV-B^{2}} \, ,
		\end{aligned}$%
	}
\end{equation}
\begin{equation}
	P = \frac{NRT}{V-B} - \frac{A}{V^{2}+2BV-B^{2}} \, ,
\end{equation}
and their derivatives with respect to mole number and volume are given as below
\begin{equation}
    \left(\frac{\partial \mu}{\partial N}\right)_{V,T} = \frac{RT V^{2}}{N(V-B)^{2}} - \frac{aV}{V^{2}+2BV-B^{2}} - \frac{aV\Big(V^{2}+B^{2}\Big)}{\Big(V^{2}+2BV-B^{2}\Big)^{2}} \, ,
\end{equation}
\begin{equation}
    \left(\frac{\partial \mu}{\partial V}\right)_{N,T} = -\left(\frac{\partial P}{\partial N}\right)_{V} = -\frac{RT V}{(V-B)^{2}} + \frac{2aNV\Big(V+B\Big)}{\Big(V^{2}+2BV-B^{2}\Big)^{2}} \, ,
\end{equation}
\begin{equation}
    \left(\frac{\partial P}{\partial V}\right)_{N,T} = -\frac{NRT}{(V-B)^{2}} + \frac{2A\Big(V+B\Big)}{\Big(V^{2}+2BV-B^{2}\Big)^{2}}\, .
\end{equation}
In the above expressions, $A = aN^{2}$, $B = bN$ and the universal gas constant $R \approx 8.3145 \; (\text{Pa}\cdot\text{m}^{3})/(\text{K}\cdot\text{mol})$. The energy parameter $a$ is defined in \ref{sec:appA} as well. In this study, we will calculate the minimum energy path and saddle point for a single-component VT flash model by solving 
\begin{equation}\label{NVTdynamic}
\begin{split}
      &\frac{\partial n_{g}}{\partial t}=-\mu(n_{g})V_{g}+\mu(n_{l})V_{g} \, , \\
      & \frac{\partial V_{g}}{\partial t}=\mu(n_{g})n_{g}-f(n_{g})-\Big(\mu(n_{l})n_{l}-f(n_{l})\Big) \, ,
\end{split}
\end{equation}
where $n_g=\dfrac{N_g}{V_g}$, $n_l=\dfrac{N_l}{V_l}$ and $f=\dfrac{F}{V}$ represents the Helmholtz free energy density.

\section{The string method}
The main purpose of using the string method is to find the minimum energy path (MEP) for barrier-crossing events. If the potential energy $F$ of a system has at least two local minima $m_1$ and $m_2$, a MEP is a curve $\gamma$ which connects these two local minima and tangents to $\nabla F$
\begin{equation}
    \nabla F(\gamma)-(\nabla F(\gamma),\hat{\gamma})\hat{\gamma}=0 \, ,
\end{equation}
where $\hat{\gamma}$ denotes the unit tangent of $\gamma$. 
Basically, finding the MEP by string is to evolve a curve connecting two local minima under some certain dynamics and the overall algorithm is an iterative application of two main procedures: using ordinary differential equation (ODE) solvers to evaluate the string and reparameterizing the string by interpolation. 

For the first step, as shown in \cite{WeinanE2007}, $N$ discrete points along the string are evolved by 
\begin{equation}\label{dy1}
    (\phi_i)_t=-\nabla F(\phi_i),\quad i=1,\ldots,N \, .
\end{equation}

For the reparametrization procedure, it is worth pointing out that only the normal component of the curve's velocity affects the curve, while the tangential velocity only contributes to the movement of points along the curve. Thus, it is free to choose any parametrization and here a particular parametrization of the curve $\Gamma=\{\phi(\alpha):\alpha\in[0,1]\}$ is used. Whenever the curve evolves, points on the $\Gamma$ will change their positions following the dynamic \eqref{dy1}. It is then necessary to interpolate these points and their values onto a same parametrization $\Gamma$. Here we give two common ways of parametrization.

\textbf{Parametrization by equal arc length}\\
In this parametrization, we have two steps:\\
(1)\quad Calculate the arc length $l_i$ based on the current point set $\{\phi_i\}$ of the curve 
\begin{equation*}
    l_i=l_{i-1}+|\phi_{i}-\phi_{i-1}|,\quad i=1,\ldots,N \, ,
\end{equation*}
where $l_0=0$ and then the nonuniform mesh $\{m_i\}$ can be obtained by normalizing $\{l_i\}$
\begin{equation*}
    m_i=l_i/l_N \, .
\end{equation*}
(2)\quad Estimate new $\phi_i^{*}$ at the uniform mesh $\{m_i^{*}=i/N\}$ under given $\{m_i\}$ and $\{\phi_i\}$ by interpolation.

\textbf{Parametrization by energy-weight arc length}
\\ It is evident that the saddle point in MEP has the largest energy. In order to give finer resolution around the saddle point, another parametrization process is carried out, which includes the following two steps:\\
(1)\quad Calculate the energy-weighted arc length $l_i$ based on the current point set $\{\phi_i\}$ of the curve 
\begin{equation*}
    l_i=l_{i-1}+W(\dfrac{F(\phi_{i-1})+F(\phi_{i})}{2})|\phi_{i}-\phi_{i-1}|,\quad i=1,\ldots,N \, ,
\end{equation*}
where $l_0=0$ and $W(x)$ is weight function which can be some well-selected positive and increasing functions for $x\in R$. Then the nonuniform mesh can be generated under the same procedure as in the first parametrization process.\\
(2)\quad Estimate new $\phi_i^{*}$ at the uniform mesh $\{m_i^{*}=i/N\}$ under given $\{m_i\}$ and $\{\phi_i\}$.
%In this work, we use exponential time differencing scheme to avoid the effects of rigid Hessian matrix of the system and this will be discussed in detail in the next section. 

\section{ Rosenbrock exponential time differencing schemes for stiff system with PR EOS}
For VT flash problems with strong stiffness \eqref{NVTdynamic}, conventional ODE solvers do not converge and have large errors. Instead, we use Rosenbrock-Euler exponential time differencing schemes to reduce the numerical difficulty caused by the stiffness of the Hessian matrix. For this purpose, first a linear term is introduced into the original system and the new system is written as
\begin{equation}\label{Newdynamic}
    \frac{du}{dt}=H(u_n)u+G(u,u_n) \, ,
\end{equation}
where $H(u_n)=H_n$ is the Hessian matrix of $F(u_n)$ , $u_n$ is the molar density at $n_{th}$ time step and
$$
    G(u,u_n)=-\nabla F(u)-H(u_n)u \, .
$$
From \eqref{Newdynamic}, it gives
\begin{equation}\label{ee}
(e^{-tH_n}u)_t=e^{-tH_n}G(u,u_n) \, .
\end{equation}
Then we integrate \eqref{ee} on both sides from $0$ to one time step $\tau$, which yields
\begin{equation}\label{ETDqian}
    \begin{split}
        &e^{t_{n+1}H_n}u_{n+1}- e^{-t_{n}H_n}u_{n}=\int_0^\tau e^{(t_n+s)H_n}G(u(t_n+s),u_n)ds \, ,\\
        &u_{n+1}=e^{\tau H_n}u_n+\int_0^\tau e^{(\tau-s)H_n}G(u(t_n+s),u_n)ds \, .
    \end{split}
\end{equation}
We can see that, different from ordinary ODE solvers, the Rosenbrock ETD scheme gives the exact relationship between $u_{n+1}$ and $u_n$. To some extent, the influence of Hessian matrix on the convergence performance is negligible and thus it allows us to avoid using the modified Cholesky algorithm that is extensively used in the literature. For the term $u(t_n+s)$ in \eqref{ETDqian}, if we use different numerical methods to approximate it, we can get different ETD schemes for the system \eqref{NVTdynamic}. For instance, by replacing $u(t_n+s)$ with $u_n$, the following first-order ETD scheme (ETD1) is established
\begin{equation}\label{ETD1}
    u_{n+1}=e^{\tau H_n}u_n+H_n^{-1}(e^{\tau H_n}-I)G(u_n,u_n) \, ,
\end{equation}
where $I$ is the identity matrix.
% Also, in some cases, we need high accurate solution to reduce the difficulty of numerical calculation caused by stiff system
However, in some cases, the high-accuracy solution is needed to reduce the difficulty of numerical calculation caused by stiff system. Here we introduce a second-order ETD (ETD2) scheme in which $u(t_n+s)$ is approximated by linear interpolation
\begin{equation*}
    \begin{split}
        \tilde{u_{n+1}}&=e^{\tau H_n}u_n+H_n^{-1}(e^{\tau H_n}-I)G(u_nH_n,u_n) \, ,\\
        u_{n+1}&=e^{\tau H_n}u_{n}+\int_0^{\tau}e^{(\tau-s)H_n}[(1-\frac{s}{\tau})G(u_n,u_n)+\frac{s}{\tau}G(\tilde{u_{n+1}},u_n)]ds \, ,\\
        &=e^{\tau H_n}u_n+\tau[(\phi_0(\tau H_n)-\phi_1(\tau H_n))G(u_n,u_n)+\phi(\tau H_n)H(u_{n+1},u_n)] \, ,
    \end{split}
\end{equation*}
where
\begin{equation*}
    \begin{split}
        &\phi_0(\alpha)=(e^{\alpha}-I)\alpha^{-1} \, ,\\
        &\phi_1(\alpha)=(e^{\alpha}-I-\alpha)(\alpha^2)^{-1} \, .
    \end{split}
\end{equation*}

\begin{table}[H]
    \caption{Compositional parameters of the investigated species in numerical examples.}
    \label{Tab1}
    \makebox[\linewidth]{
    	\begin{tabular}{ c c c c c }
    	    \hline
    		Component & $T_{\mathrm{c},i} \, (\mathrm{K})$ & $P_{\mathrm{c},i} \, (\mathrm{MPa})$ & $M_{\mathrm{w},i} \, (\mathrm{g}/\mathrm{mol})$ & $\omega_{i}$ \\
    		\hline
    		$\mathrm{nC}_{4}$ & 425.12 & 3.796 & 58.12 & 0.2010 \\
    		$\mathrm{CO}_{2}$ & 304.14 & 7.375 & 44.01 & 0.2390 \\
    		\hline
    	\end{tabular}
    }
\end{table}

\section{Numerical experiment}
In this section, we calculate the MEP and the saddle-point for the single-component VT flash model. 
We first use Rosenbrock-Euler-ETD scheme to solve the single-component VT flash problem. The numerical results will be compared with both published results and experimental data to demonstrate the feasibility of the Rosenbrock-Euler-ETD scheme in dealing with the VT flash problem with strong stiffness.

\subsection{Equilibrium calculation with Rosenbrock-Euler-ETD scheme}
The performance of the Rosenbrock-Euler-ETD scheme on the VT flash model using PR EOS is tested for both hydrocarbon component and non-hydrocarbon component. Compositional parameters of the investigated species are shown in Table \ref{Tab1}.
To confirm the accuracy and credibility of the proposed algorithm, the computed results are compared with the experimental data, which are obtained from NIST chemistry webbook \cite{NIST}. Numerical experiments are conducted in two ways: A) fixing the total mole number $N$ and changing the temperature $T$; B) fixing the temperature $T$ and changing the total mole number $N$. The total volume $V$ in both cases are same and assumed to be unity.

\begin{figure}[ht!]
    \centering
    \includegraphics[width=3.6in,height=2.6in]{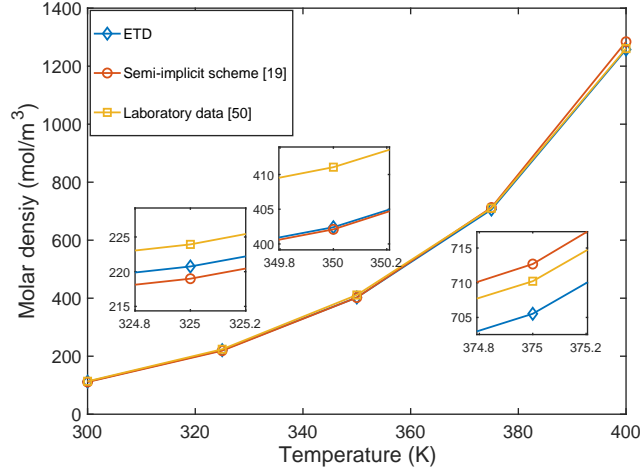}
    \caption{Equilibrium molar density of the gas phase at $T = 300, 325,350, 375, 400 $ K and the comparison with published results and experimental data ($\text{nC}_{4}$)}
    \label{fig1}
\end{figure}

In the numerical experiment of hydrocarbon component, the investigated species is n-butane ($\text{nC}_{4}$). For the experiment A, the total particle number is $N = 3000$ mol. Temperatures in the numerical experiment are $300$, $325$, $350$, $375$ and $400$ K. The time step is $10^{-4}$. For the experiment B, the temperature is fixed to $350$ K and the total mole number $N = 2000, \, 3500, \, 5000, \, 6500, \, 8000$ mol.

\begin{figure}[ht!]
    \centering
    \includegraphics[width=3.6in,height=2.6in]{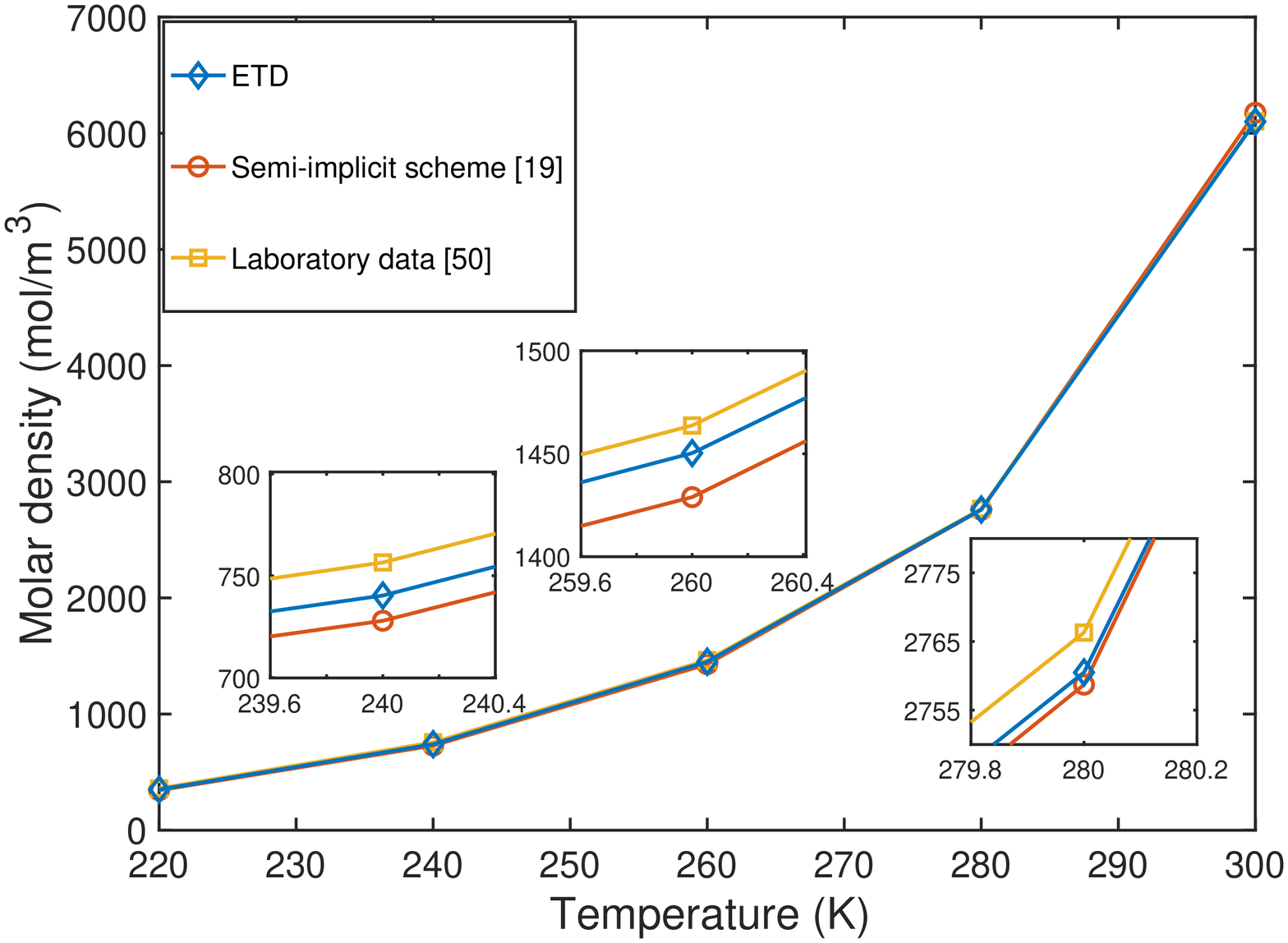}
    \caption{Equilibrium molar density of the gas phase at $T = 220, 240,260, 280, 300 $ K and the comparison with published results and experimental data ($\text{CO}_{2}$)}
    \label{fig2}
\end{figure}

In the numerical experiment of non-hydrocarbon component, the investigated substance is $\text{CO}_{2}$. For the experiment A, the total particle number is $N = 10000$ mol. Temperatures in the numerical experiment are $220$, $240$, $260$, $280$ and $300$ K. The time step is $10^{-4}$. For the experiment B, the temperature is fixed to $280$ K and the total mole number $N = 5000, \, 7500, \, 10000, \, 12500, \, 15000$ mol.

Figure \ref{fig1} and Figure \ref{fig2} presents the molar density of gaseous $\text{nC}_{4}$ and $\text{CO}_{2}$ at the equilibrium state, respectively. As temperature increases, the molar density of gas phase goes up because an increasing number of molecules evaporate into the gas phase. To test the prediction accuracy of the proposed ETD scheme, we compare the computed results with our previous work \cite{Yiteng2018}
and experimental data. The inset figures show more details which cannot be clearly seen from the parent figure.
It can be seen that the molar densities calculated by the Rosenbrock-Euler-ETD scheme agree with the experimental data very well. By comparing with the results of our previous work, the explicit scheme we used in this paper exhibits the same accuracy as the semi-implicit scheme in \cite{Yiteng2018}. 
Table \ref{Tab2} and Table \ref{Tab3} give the equilibrium mole number, volume and molar density of gaseous $\text{nC}_{4}$ and $\text{CO}_{2}$ at fixed temperatures, respectively. The constant molar densities can be explained in terms of the Gibbs phase rule. Since we fix the temperature, for a single-component system, the pressure is fixed as well. At the same isothermal-isobaric condition, the densities of gas and liquid phase keep constant so that the variation of gas volume is identically proportional to the variation of the mole number of gas phase. As we can see, our new scheme works better in the constant temperature cases than the previous work \cite{Yiteng2018}.

\begin{table}[H]
    \caption{Equilibrium mole number ($N_{g}$), volume ($V_{g}$) and molar density ($n_{g}$) of gaseous $\text{nC}_{4}$ at $T = 350$ K with total mole number $N = 2000, \, 3500, \, 5000, \, 6500, \, 8000$ mol}
    \label{Tab2}
    \makebox[\linewidth]{
    	\begin{tabular}{ c c c c c c}
    		\hline
    		$N$ &  $N_g$ (ETD)& $V_g$ (ETD)&  $n_g$ (ETD) & $n_g$  ([19]) & laboratory data\\
    		\hline
    		2000 & 326.3 & 0.812 & 402.2 & 402.1 & 402.4\\
            3500 & 254.9 & 0.634 & 402.2 & 402.1 & 402.4\\
            5000 & 184.2 & 0.458 & 402.2 & 402.1 & 402.4\\
            6500 & 113.0 & 0.281 & 402.2 & 402.1 & 402.4\\
            8000 & 41.8 & 0.104 & 402.2 & 402.1 & 402.4\\
    		\hline
    	\end{tabular}
    }
\end{table}

\begin{table}[H]
    \caption{Equilibrium mole number ($N_{g}$), volume ($V_{g}$) and molar density ($n_{g}$) of gaseous $\text{CO}_{2}$ at $T = 280$ K with total mole number $N = 5000, \, 7500, \, 10000, \, 12500, \, 15000$ mol}
    \label{Tab3}
    \makebox[\linewidth]{
    	\begin{tabular}{ c c c c c c}
    		\hline
    	    $N$ &	$N_g$ (ETD)& $V_g$ (ETD)& $n_g$ (ETD)& $n_g$  ([19]) & laboratory data\\
    	    \hline
    		5000 & 2388.4 & 0.865 & 2763.3 & 2758.7 & 2766.3\\
            7500 & 1973.7 & 0.715 & 2763.3 & 2758.7 & 2766.3\\
            10000 & 1559.1 & 0.565 & 2763.3 & 2758.7 & 2766.3\\
            12500 & 1144.4 & 0.415 & 2763.3 & 2758.7 & 2766.3\\
            15000 & 729.7 & 0.2644 & 2763.3 & 2758.7 & 2766.3\\
    		\hline
    	\end{tabular}
    }
\end{table}

% The hydrocarbon component used in numerical experiments is n-butane ($\text{nC}_{4}$). Then, an approximation of MEP is identified  by using the string method with $N=100$ images under temperature 350K and total particle number $N=3000$.
\begin{figure}[ht!]
    \centering
    \includegraphics[width=3.6in,height=2.6in]{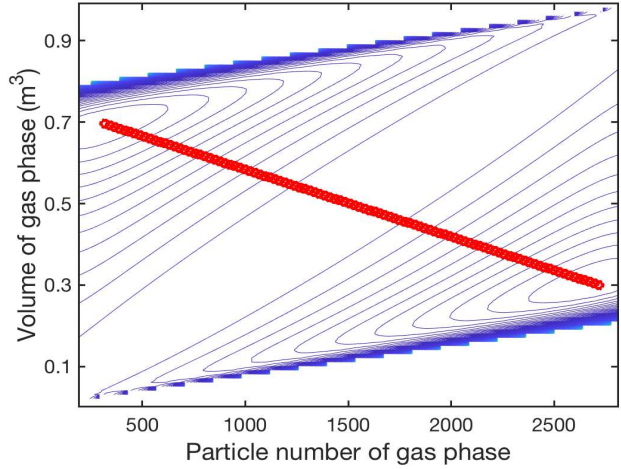}
    \caption{Initial configuration of string in the energy landscape for $\text{nC}_{4}$}
    \label{fig3}
\end{figure}

\subsection{MEP of a single-component VT flash model with PR EOS}
In this section, an example will be presented to calculate the MEP and one of the saddle point of the VT flash model for pure $\text{nC}_{4}$ at temperature $T = 350$ K. The total particle number is $3000$ mol and the time step is $10^{-6}$. During the calculation process, the MEP was approximated using the string method with 100 images and the initial string is the linear interpolation between $(316, \, 0.69)$ and $(2700, \, 0.3)$, see Figure \ref{fig3}. The two end points of initial string are not required close to the local minima. For each point on the string, a gradient flow calculation is performed in each time step. These two end points of the string will converge to local minima. Cubic spline interpolation is used to redistribute the points of the string in each time step according to equal arc length. The whole process is evolving until
\begin{equation*}
    \max\dfrac{1}{\Delta t}|u_i^{n+1}-u_i^n|<\text{TOL} \, ,
\end{equation*}
where the tolerance $\text{TOL}=10^{-8}$.

\begin{figure}[ht!]
    \centering
    \includegraphics[width=3.6in,height=2.6in]{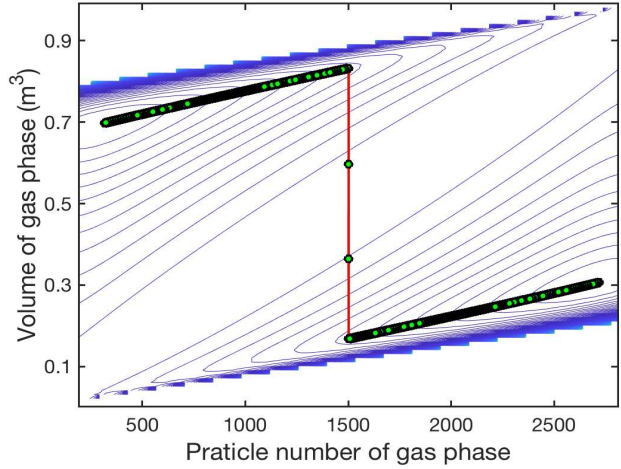}
    \caption{Minimum energy path of $\text{nC}_{4}$}
    \label{fig4}
\end{figure}

The calculated MEP is shown in Figure \ref{fig4}. 
%If we change the reparameterized mode by energy-weight arc length, the first order saddle point $(1500, \, 0.5)$ of the energy landscape can be calculated.
If we use the two-step procedure introduced in \cite{WeinanE2007}, We can determine the location of the saddle point $(1500, \, 0.5)$ of the energy landscape by using the following climbing image algorithm
\begin{equation}\label{climb}
    \phi^{\prime}_{t}=-\nabla F(\phi^{\prime})+2(\nabla F(\phi^{\prime}),\hat{\gamma_{0}^{\prime}})\hat{\gamma_{0}^{\prime}},
\end{equation}
where $\phi^{\prime}$ is the saddle point of $\nabla F=0$ and $\gamma^{\prime}$ is the related unit tangent vector of the MEP. It can be proved that the stable solution of (\ref{climb}) is the unstable solution of $\nabla F=0$. The initial value of the algorithm requires that the initial $\phi^{\prime}_{0}$ is close to the actual saddle point. Here, we use the string method to calculate a rough approximation of MEP with small number of images (in this work, we choose 25 images). Then $\phi^{\prime}_{0}$ can be selected as the point with the highest potential energy on the MEP and $\gamma_{0}^{\prime}$ is the related unit tangent vector.

\section{Conclusion}
In this study, we have applied the string method to calculate the minimum energy path of the single-component VT flash model with the Peng-Robinson equation of state. Especially, in order to avoid the computational difficulties brought by strong stiffness of the model, the Rosenbrock-type ETD scheme, which is believed superior to other common ODE solvers, is used to evolve images of the string in each time step. Numerical experiments are carried out and show good feasibility and accuracy of the proposed algorithm by comparing the computed results with the results of previous work and experimental data. It is worth pointing out that the local minima has been connected by a MEP of the given model and consequently its saddle point can be accurately calculated. This work has discussed the relation between the local minima of a single-component VT flash system and its saddle point for the first time. However, it is still a difficult task either to get the high-order saddle points or to find the global minimum energy point for a multi-component VT flash system, which is the subject of our future work.

\section*{Acknowledgments}
The authors greatly thank for the support from the National Natural Science Foundation of China (grant number 51874262, % 51904031
51936001,11802090) and the Research Funding from King Abdullah University of Science and Technology (KAUST) through the grants BAS/1/1351-01, URF/1/4074-01, URF/1/3769-01, and REP/1/2879-01.



\appendix
\section{Parameters of the Peng-Robinson equation of state} \label{sec:appA}
The Peng-Robinson equation of state has the following form
\begin{equation}
    P(\bm{N}, V, T) = \frac{NRT}{V-bN} - \frac{aN^{2}}{V^{2}+2bNV-(bN)^{2}} \, ,
\end{equation}
and accordingly the Helmholtz free energy can be defined as
% \begin{equation}
%     \begin{split}
%       & f(\mathbf{n})=f^{\text{ideal}}(\mathbf{n})+f^{\text{excess}}(\mathbf{n}),\\
%       & f^{\text{ideal}}(\mathbf{n})=RT\sum_{i=1}^Mn_i(\text{ln}n_i-1),\\
%       & f^{\text{excess}}(\mathbf{n})=-nRT\text{ln}(1-bn)+\frac{a(T)n}{2\sqrt{2}b}\text{ln}(\frac{1+(1-\sqrt{2})bn}{1+(1+\sqrt{2})bn}).
%     \end{split}
% \end{equation}
% where $\mathbf{n}=(n_1,...,n_M)$, $n=\sum_{i=1}^Mn_i$ and $n_i=\dfrac{N_i}{V}$.
\begin{equation}
    \resizebox{\textwidth}{!}{%
		$\begin{aligned}
			F(\bm{N}, V, T) = RT\sum_{i=1}^{M}N_{i}\left(\ln\frac{N_{i}}{V}-1\right) - NRT\ln\left(\frac{V-B}{V}\right) + \frac{A}{2\sqrt{2}B}\ln\left(\frac{V+(1-\sqrt{2})B}{V+(1+\sqrt{2})B}\right) \, ,
		\end{aligned}$%
	}
\end{equation}
where the coefficients $A = aN^{2}$ and $B = bN$, and $N = \sum_{i}N_{i}$ is the total mole number.
The energy parameter $a$ and co-volume parameter $b$ of a fluid mixture are computed based on the classical Van der Waals mixing rule such that
\begin{equation}
    a = \sum_{i}\sum_{j}x_{i}x_{j}\left(a_{i}a_{j}\right)^{1/2}\left(1-k_{ij}\right) \, , \quad b = \sum_{i}x_{i}b_{i} \, .
\end{equation}
In the above expression, $k_{ij}$ is the binary interaction coefficient. For pure-component fluid systems, $x_{i}=1$ and $k_{ij}=0$ so that $a$ and $b$ are reduced to
\begin{equation}
    a = a_{i} = 0.45724\frac{R^{2}T_{c}^{2}}{P_{c}}\left(1+m \left(1-\sqrt{\frac{T}{T_{c}}}\right)\right)^{2} \, , \\
\end{equation}
\begin{equation}
    b = b_{i} = 0.07780\frac{RT_{c}}{P_{c}} \, .
\end{equation}
Here, $T_{c}$ and $P_{c}$ represent the critical temperature and the critical pressure of a pure substance, respectively. The parameter $m$ is a function of the acentric factor $\omega$ of pure substance
\begin{equation}
    m = 0.37464 + 1.54226\omega - 0.26992{\omega^2} \, ,\quad \omega \le 0.49 \, ,
\end{equation}
\begin{equation}
    m = 0.379642 + 1.485030\omega  - 0.164423{\omega^2} + 0.016666{\omega^3} \, , \quad \omega > 0.49 \, .
\end{equation}
If $\omega$ is unavailable, it can be estimated by the following correlation
\begin{equation}
    \displaystyle {\omega} = \frac{3}{7}\left(\frac{{{{\log }_{10}}\left(\displaystyle \frac{{P_{c}}}{{1 \; {\rm{atm}}}}\right)}}{\displaystyle {\frac{{T_{c}}}{{T_{b}}} - 1}}\right) - 1 \, ,
\end{equation}
where $T_{b}$ is the normal boiling point.\\ 
% The definition of parameter $\kappa$ is related to the P-R parameters $a$ and $b$ by
% $$
% \kappa = a {b^{2/3}}(m_1^c(1 - \frac{T}{{{T_{c}}}}) + m_2^c),
% $$
% here, $m_1^c$ and $m_2^c$ denote the coefficients which can be related to the acentric factor $\omega$ by
% $$
% m_1^c =  - \frac{{{{10}^{ - 16}}}}{{1.2326 + 1.3757\omega }}, \quad m_2^c = \frac{{{{10}^{ - 16}}}}{{0.9051 + 1.5410\omega }}.
% $$

%%       \citet{<label>} ==> Jones et al. [21]
%%       \citep{<label>} ==> [21]
%%






% \end{thebibliography}

\end{document}